\documentstyle[preprint,aps]{revtex}
\begin{document}
\preprint{Revised on \today}
\title{Impact of Charge Ordering on Magnetic Correlations
in Perovskite (Bi,Ca)MnO$_3$}
\author{Wei Bao,$^{1}$ J. D. Axe,$^{1}$
C. H. Chen$^{2}$ and S-W. Cheong$^{2}$}
\address{$^{1}$Brookhaven National Laboratory, Upton, NY 11973\\
$^{2}$Bell Laboratories, Lucent Technologies, Murray Hill, NJ 07974}
\maketitle

\begin{abstract} 
Single crystalline Bi$_{1-x}$Ca$_{x}$MnO$_3$ ($0.74 \leq x \leq  0.82$)
were studied with neutron scattering, electron diffraction
and bulk magnetic measurement. We discovered dynamic ferromagnetic spin
correlations at high temperatures, which are replaced by antiferromagnetic
spin fluctuations at a concomitant charge ordering and structural transition. 
Our results indicate that thermal-activated hopping of the Jahn-Teller
active $e_g$ electrons in these insulating materials, nevertheless, induce ferromagnetic interaction through double-exchange mechanism. It is
the ordering of these charges competing with the double-exchange
ferromagnetic metallic state.
\end{abstract}

\narrowtext

The relation between double exchange and
Jahn-Teller distortion is one of the central issues\cite{de_latt}
in current research on perovskite mixed-valent 
manganites $T_{1-x}D_x$MnO$_3$ 
($T$ being trivalent ions, e.g., La, Pr or Nd, and $D$ being
divalent ions, e.g., Ca, Sr, Ba or Pb\cite{rev_g}).
In these materials, the mean number of $3d$ electrons per Mn 
can be tuned from 4 to 3.
Due to strong Hund rule coupling, Mn$^{3+}$ has a high-spin 
state $t_{2g}^3 e_g^1$ 
and the $e_g$ level is split by Jahn-Teller distortion. 
For the Mn$^{3+}$-end members, the long range order of the elongated MnO$_6$
octahedra (with $d_{z^2}$-type orbitals for this kind of Jahn-Teller distortion)
stabilizes an insulating antiferromagnetic
ground state\cite{wollan,orb_ge,orb_fuji}.
In doped samples, $e_g$ electron hopping\cite{hopp} 
between Mn$^{3+}$ and Mn$^{4+}$ neighbor
induces ferromagnetic correlations through
the so-called double exchange mechanism\cite{de_zener}.
It is well known that for $x$ larger than a threshold value,
a ferromagnetic metallic ground state becomes stable\cite{jonker,park}. 
A structural transition to a higher symmetry state also occurs
at the threshold\cite{LaCa-end}, which results from the ability of the
hopping $e_g$ electrons to destroy the
long range coherence of the Jahn-Teller distortion (or 
$d_{z^2}$-type orbitals)\cite{orb_ge}.
Long range order of charges in the Mn$^{4+}$($t^3_{2g}$)-rich regime 
was recently observed with electron diffraction 
in La$_{1-x}$Ca$_x$MnO$_3$ ($x>0.5$)\cite{bao95b}.
A large anomaly in sound velocity was found at the charge ordering temperature. 
More interestingly,
the magnetic susceptibility $\chi$(T) has a pronounced inflection at 
the charge ordering temperature, resembling that associated with a
conventional long range antiferromagnetic transition\cite{bao94c}. 
However, thermodynamic measurements suggest that the antiferromagnetic order 
develops at a lower temperature\cite{bao95b}.
In order to clarify and understand the anomalous lattice and magnetic
phenomena at charge ordering, microscopic
information is crucial, and neutron scattering is the ideal probe
for this purpose with its access to static and dynamic correlations of
both magnetic and structural degrees of freedom\cite{theo}.

Single crystals of La$_{1-x}$Ca$_x$MnO$_3$ ($x>0.5$) are not available. 
However, we have succeeded in growing single crystals of isostructural 
Bi$_{1-x}$Ca$_x$MnO$_3$ in this Mn$^{4+}$-rich regime. 
As the following description will show, (Bi,Ca)MnO$_3$ bears 
close resemblance to (La,Ca)MnO$_3$ in the 
overlapping composition range,
attesting to the identical underlying physics.
Single crystals of (Bi,Ca)MnO$_3$, however, allow us to extract much 
more information with neutron scattering. We found that the charge 
ordering is accompanied by a structural transition and that antiferromagnetic
long range order indeed develops at a lower temperature. The most important 
of our discoveries, however,  is that 
the nature of spin fluctuations change from {\em ferro}magnetic to
{\em antiferro}magnetic at the charge ordering transition.
These results indicate that at high temperatures, thermally
activated $e_g$ electron hopping
in the insulating regime, nevertheless, induces ferromagnetic correlations
through the double-exchange mechanism. When $e_g$ electrons with 
their accompanying Jahn-Teller distortions freeze in static order, 
the double exchange induced ferromagnetic spin fluctuations are 
replaced by superexchange antiferromagnetic spin fluctuations.

Single crystals of Bi$_{1-x}$Ca$_x$MnO$_3$ were grown using the flux method. 
The compositions of samples were determined by inductive coupled
plasma emission spectroscopy.
The weights of samples used in neutron scattering are 25, 127 and 320mg 
for $x$=0.74, 0.76 and 0.82, respectively.
Charge ordering was observed with electron diffraction using
a JOEL 2000FX transmission electron microscope. 
Superlattice peaks with wave vector ($\delta,0,0$) in $Pbnm$ notation
appear below a charge ordering temperature, T$_O$, similar to those 
in (La,Ca)MnO$_3$\cite{bao95b}. $\delta\simeq 0.30$ and 0.22 for
$x=0.74$ and 0.82, respectively. The decreasing $\delta$
is consistent with the decreasing numbers of $e_g$ electrons, which
are involved in the charge ordering.
Fig.~1(a) shows the development of the charge superlattice intensity 
of (0.22,0,0)
below T$_O =210$K for a $x$=0.82 sample.
Magnetic susceptibility was measured for crystals from the same batches with
a Quantum Design SQUID magnetometer. Consistent results were found for different
crystals from the same batch. Fig.~1(b) shows an example for $x$=0.82.
At high temperatures, the Curie-Weiss law is followed, yielding an
effective moment $p=4.15(2)\mu_B$ which is close to
the expected $p=4.08$ for this composition.
The positive value of the Weiss constant $\Theta=159(1)$K reveals
the existence of ferromagnetic correlations between Mn spins.
The drastic reduction of $\chi$(T) at the charge ordering
temperature, T$_O$, is similar to 
that observed in (La,Ca)MnO$_3$\cite{bao94c}.
At 160K, there begins a further reduction in the magnetic susceptibility, 
whose origin will become
clear after the presentation of our neutron scattering results.

Neutron scattering experiments were conducted with triple
axis spectrometers at the HFBR at Brookhaven National Laboratory.
Except for polarized neutron scattering,
graphite monochromators and analyzers were used. High order 
contaminations were removed with graphite filters.
At room temperature, the selection rules for nuclear Bragg peaks 
are consistent with space group $Pbnm$. 
However, the orthorhombic distortion is small.
For the purpose of this paper, we use the simpler pseudocubic 
perovskite unit cell\cite{qzhu}.
Fig.~1(c) shows the pseudocubic lattice parameters as a function of 
temperature for a $x$=0.82 sample.
At the charge ordering temperature T$_O$, there is a structural transition, 
as evidenced by the splitting of the Bragg peaks.
This result is not totally surprising, since
each $e_g$ electron carries with it a Jahn-Teller lattice distortion.
When $e_g$ electrons order, long range correlations can develop among
Jahn-Teller distorted MnO$_6$ octahedra\cite{orb_ge}.

In a pioneering powder neutron diffraction study,
three types of magnetic structures were reported 
in the composition range $0.74\leq x\leq 0.82$ for 
La$_{1-x}$Ca$_x$MnO$_3$\cite{wollan}.
We have conducted thorough search in (hk0) and (hhl) zones 
of (Bi,Ca)MnO$_3$ at low temperatures. 
For all of our samples, magnetic Bragg peaks are found
only at C-type points, ($\frac{2n+1}{2},\frac{2m+1}{2},l$).
We distinguished magnetic peaks from  structural superlattice peaks not
only with temperature dependence, but more conclusively using 
polarized neutron scattering. Heusler crystals were used both as
monochromator and analyzer, and a spin flipper and a graphite filter
were placed in the diffracted beam. 
The inset on the right of Fig.~1(d) shows an example of a (1/2,1/2,0)
scan for the $x$=0.82 sample.
By aligning a magnetic field along different directions at the sample
position, the magnetic nature of the Bragg peak is  proven and
the spin direction is determined\cite{moonr}. In the current case,
spin lies along the c-axis, as found 
in (La,Ca)MnO$_3$ for a C-type antiferromagnet\cite{wollan}.
The resulting spin structure is shown in the left insert to Fig.~1(d). 
Possible implications of this magnetic structure for $e_g$ orbital ordering
has been discussed by Goodenough\cite{orb_ge}.
Comparing the magnetic and nuclear Bragg intensities,
the staggered magnetic moment at 9K is determined to be 3.5(2)$\mu_B$
per Mn. This corresponds to an average Mn spin $S=1.8(1)$ 
and $p=2\surd \overline{S(S+1)}=4.4(3)$ is in good agreement with 
that given by susceptibility measurement at high temperatures.
The temperature dependence of the order parameter in
Fig.~1(d) establishes the N\'{e}el temperature T$_N=160$K for $x=0.82$ 
which coincides with the lower inflection point of $\chi(T)$ in Fig.~1(b).
The N\'{e}el temperature T$_N$ and the charge-structural transition 
temperature T$_O$ for various samples are summarized in the insert to
Fig.~1(a).
This phase diagram parallels that of (La,Ca)MnO$_3$\cite{bao95b},
pointing to a generic behavior for both systems.

After establishing the phase relation and the magnetic ground state, 
let us now turn to dynamic spin properties. 
Fig.~\ref{disp}(a) shows spin-wave dispersion along three symmetry directions
for $x=0.82$ measured at 11K near a magnetic zone center.
The spin Hamiltonian which describes Mn spins below charge ordering transition is
\[
{\cal H}=\sum_{\langle i,j \rangle_{AF}} J_{AF} {\bf S}_i \cdot {\bf S}_j
+\sum_{\langle i,j \rangle_{F}} J_{F} {\bf S}_i \cdot {\bf S}_j
-g \mu_B H_u \sum_i |S^z_i|
\]
where $J_{AF}$ and $J_F$ are antiferromagnetic 
and ferromagnetic nearest neighbors interactions
(refer to Fig.~\ref{stat}(d))
and $H_u$ is a uniaxial anisotropy field.
Each nearest neighbor spin pair is counted only once in the summations.
Conventional spin wave theory yields the dispersion relation\cite{theo}
\begin{equation}
\hbar \omega({\bf q})=\surd \overline{[{\cal J}(0)
-{\cal J}_1(0)+{\cal J}_1({\bf q})+g\mu_B H_u]^2
-{\cal J}({\bf q})^2}
\end{equation}
where ${\bf q}=(hkl)$ is the wave vector from an antiferromagnetic zone
center,
$ {\cal J}({\bf q})=2S J_{AF} \bigl(\cos(2\pi h)+\cos(2\pi k)\bigr)$
and
$ {\cal J}_1({\bf q})=2S J_{F} \cos(2\pi l)$.
This dispersion relation accounts well for our data, 
as indicated by the solid curves in Fig.~\ref{disp}(a).
From the fit we derive $S J_{AF}=3.6(1)$meV, $S J_F=-1.3(1)$meV 
and $H_u=4.4(3)$T. The zone boundary energy 
at (0,0,1/2) derived using these parameters 
is 13.8(8)meV, which corresponds nicely to the observed $k_B$T$_N$
as expected for a bi-partite 3-dimensional magnet.
Recently, the spin wave dispersion has been measured for 
ferromagnetic samples\cite{sw_La-Sr,sw_La-Pb,sw_La-Ca}. 
A Heisenberg nearest neighbor model describes well the
experimental results at low temperatures. 
The nearest neighbor exchange varies from $SJ=-12.6(5)$meV at 27K
to $-7.6(3)$meV at 300K for La$_{0.7}$Sr$_{0.3}$MnO$_3$\cite{sw_La-Sr};
for La$_{0.7}$Pb$_{0.3}$MnO$_3$ and La$_{0.77}$Ca$_{0.33}$MnO$_3$, 
the corresponding value at T$\rightarrow$0 is $-8.8(2)$meV\cite{sw_La-Pb} 
and $-11.4$meV\cite{sw_La-Ca}, respectively. 
Ferromagnetic double-exchange in these samples appears to be stronger
than the antiferromagnetic superexchange measured in our sample\cite{hirota}.

There is an energy gap $\Delta = 3.8$meV in the spin wave 
excitations of Fig.~2(a). Compared with $k_B$T$_N=13.8$meV, this is a very
large spin gap.
Similar spin gaps were also found for $x=0.74$ and 0.76 samples.
A constant {\bf q} scan for $x=0.76$ at (1/2,1/2,0) is shown 
in Fig.~\ref{disp}(b).
This explains the pronounced reduction of $\chi(T)$ below T$_N$.
Similar, but less pronounced, reduction of $\chi$
in the case of La$_{1-x}$Ca$_x$MnO$_3$ ($x>0.5$)\cite{bao95b,bao94c} 
is likely also to be caused by the development of a spin gap below T$_N$. 
The softening of the spin gap with elevated temperatures was measured
and the results are shown in Fig.~\ref{disp}(b).
We explain the spin gap in our insulating samples with uniaxial
anisotropy. This could be a natural consequence of the ordering of the
$d_{z^2}$ orbitals. 
Meanwhile rapid hopping of $d_{z^2}$ electrons in the metallic phase 
renders spin space isotropic and 
no gap in spin excitations is expected for the ferromagnetic manganites.
This is indeed what was found experimentally for La$_{0.77}$Ca$_{0.33}$MnO$_3$
with neutron scattering\cite{sw_La-Ca}.

To investigate the anomalous reduction of $\chi$(T) at
the charge-structural transition,
we have directly measured dynamic spin correlations with inelastic
neutron scattering. Antiferromagnetic spin fluctuations
were probed at (1/2,1/2,0) (Fig.~3(b)) and ferromagnetic
spin fluctuations near the forward direction (Fig.~3(c)),
where magnetic form factor maximizes and structure factor
for structural fluctuations diminishes as $q^2$.
As charge order parameter grows below T$_O$ (refer to Fig~1(a)),
{\em ferromagnetic} spin fluctuations are replaced by {\em antiferromagnetic}
spin fluctuations. This may be better seen in Fig.~3(a) which
shows detail temperature variations of both the ferromagnetic
response (circles, measured at $\hbar \omega=1$meV and $q=0.17 \AA$) 
and the antiferromagnetic response (disks, measured at $\hbar \omega=1.7$meV 
and {\bf q} = (1/2,1/2,0)).
The absence of antiferromagnetic spin correlations above T$_O$
is evidenced by a flat background both in the energy scan 
and in {\bf q} scans (refer to Fig.~3(b)) at (1/2,1/2,0).
The lower energy limit of this flat background was pushed down to
0.4meV with better resolution of cold neutron scattering.
There is obviously no any ferromagnetic component in the low temperature
magnetic order (see Fig.~1(b)). This justifies our use of
low temperature data as the background for ferromagnetic response
in Fig~3(a), and insert in (c) shows the net ferromagnetic intensity
at 215K by subtracting background measured at 70K. 
Therefore our experimental results establish
that ferromagnetic interactions exist between spins above T$_O$
and they are replaced by antiferromagnetic interactions when 
charge orders.
This switching of magnetic correlations from a ferromagnetic type
to an antiferromagnetic type explains the inflection of $\chi$(T)
at T$_O$.

A separate antiferromagnetic transition below a charge ordering
transition was discovered previously in laminar perovskites
Sr$_{2-x}$La$_{x}$MnO$_4$ for $x\sim 0.5$\cite{bao95a}. 
This material also has an insulating antiferromagnetic ground state, and 
the high temperature ferromagnetic susceptibility is interrupted by
charge ordering, as in the Mn$^{4+}$-rich (La,Ca)MnO$_3$ and 
(Bi,Ca)MnO$_3$.
Meanwhile, double exchange ferromagnetic ground state coincides with
metallic conduction both in perovskite manganites  for 
$x\sim 0.3$\cite{Nd-Pb} and double layered 
(La$_{0.4}$Sr$_{0.6}$)$_3$Mn$_2$O$_7$\cite{two_l}.
This suggests that the competition between double-exchange and the 
{\em ordering}
of the Jahn-Teller active $e_g$ electron is the dominant factor in
determining the ground state of perovskite and related manganites.

In summary, we have observed directly in the Mn$^{4+}$-rich (Bi,Ca)MnO$_3$ 
a simultaneous charge ordering and structural transition.
Above the charge-structural transition temperature T$_O$, 
there exist ferromagnetic spin fluctuations, indicative of
double exchange induced by the $e_g$ electron hopping.
Below T$_O$, we discover the replacement of the ferromagnetic
correlations by antiferromagnetic spin fluctuations, 
which support the notion that the ferromagnetic 
double-exchange diminishes as charge orders.
The antiferromagnetic transition-like inflection point in
$\chi$(T) at T$_O$
is accounted for by this switch from ferromagnetic spin fluctuations to 
antiferromagnetic fluctuations. C-type long range antiferromagnetic
order was found to develop in a separate phase transition at a lower 
temperature with large uniaxial anisotropy, 
which open a gap in spin excitations.

It is a pleasure to thank C. Broholm, S. M. Shapiro, G. Shirane, Q. Zhu, 
H. Y. Hwang, B. Batlogg, A. Zheludev, M. Martin and G. Y. Xu for 
discussions and assistances.
WB acknowledges the Aspen Center for Physics where part of the work
was performed.
Work at BNL was supported by DOE under Contract No.\ DE-AC02-76CH00016.

\begin{figure}
\caption{Temperature dependences of (a) intensity of the charge
superlattice peak, (b) magnetic susceptibility $\chi$,
(c) pseudocubic lattice parameters, and (d) intensity of magnetic Bragg
peak (1/2,1/2,0) for a $x=0.82$ sample. $1/ \chi$ (circles) is also shown
in (b) with scale to the right and the straight line is a fit to the
Curie-Weiss law. 
Polarized neutron scattering of
(1/2,1/2,0) is shown in the right insert of (d): dots and circles
are for the flipper turned off (Voff) and on (Von) cases with a magnetic
field along (001); diamonds and squares for flipper off (Hoff) and on (Hon)
with the magnetic field along the scattering wave vector.
The spin structure determined is shown in the left corner of (d).
Insert in (a): temperature-composition phase diagram for
Bi$_{1-x}$Ca$_x$MnO$_3$, showing the charge-structural transition T$_O$
(upper curve) and the N\'{e}el temperature T$_N$ (lower curve).  }
\label{stat}
\end{figure}

\begin{figure}
\caption{(a)
Spin-wave dispersion along (110), (100) and (001) around 
the magnetic zone center (1/2,1/2,0), measured at 11K for $x=0.82$.
The solid curves represent a fit to the spin-wave dispersion
relation of Eq.~(1). 
(b)
Energy scans at (1/2,1/2,0) for $x=0.76$ at various temperatures.
The energy gap, $\Delta$, of the spin excitations collapses when 
the N\'{e}el T$_N=130$K is approached.  }
\label{disp}
\end{figure}

\begin{figure}
\caption{(a) Temperature dependence of antiferromagnetic response 
measured at (1/2,1/2,0) and 1.7 meV (disks), 
and ferromagnetic response at $|{\bf q}|=0.17\AA^{-1}$ and 1 meV (circles)
(marked with dashed lines in (b) and (c) respectively). 
The dotted line indicates background (see text for details).
Below the charge ordering transition at T$_O=210$K, ferromagnetic
response is replaced by antiferromagnetic response (disks).
(b) Energy scans at the antiferromagnetic zone center
(1/2,1/2,0) at 165K (squares), 190K (disks), 220K (triangles)
and 300K (diamonds). The solid line connects data at 300K which are
indistinquishable from those at 220K. Insert shows constant
$\hbar\omega=1.7$meV scans at temperatures indicated.
(c) Energy scans near the forward direction with
$|{\bf q}|=0.17\AA^{-1}$, probing ferromagnetic fluctuations, 
at 70K (squares), 215K (circles) and 300K (triangles).
Insert shows 215K data with background at 70K subtracted, and the
solid line is a Lorentzian.
}
\label{fluxx}
\end{figure}


\begin{references}


	

\bibitem{de_latt}
{ A. J. Millis, B. I. Shraiman and P. B. Littlewood, Phys. Rev. Lett. {\bf 74},
  5144 (1995)}.

\bibitem{rev_g}
{ J. B. Goodenough and J. M. Longon, in Landolt-B\"{o}rnstein, New Series III,
  {\bf 4}, Pt a (Springer-Verlag, Berlin, 1970)}.

\bibitem{wollan}
{ E. O. Wollan and W. C. Koehler, Phys. Rev. {\bf 100}, 545 (1955)}.

\bibitem{orb_ge}
{ J. B. Goodenough, Phys. Rev. {\bf 100}, 564 (1955)}.

\bibitem{orb_fuji}
{ T. Mizokawa and A. Fujimori, Phys. Rev. B {\bf 51}, 12880 (1995);
T. Saitoh et al., ibid. 13942 (1995)}.

\bibitem{hopp}
{Hopping of an $e_g$ electron inevitably carries a Jahn-Teller lattice distortion
  with it. In other words, the charge motion is polaronic. We freely use $e_g$
  electron for this Jahn-Teller polaron in this paper.}

\bibitem{de_zener}
{ C. Zener, Phys. Rev. {\bf 82}, 403 (1951);
 P. W. Anderson and H. Hasegawa, ibid. {\bf 100}, 675 (1955);
 P.-G. de Gennes, ibid. {\bf 118}, 141 (1960)}.


\bibitem{jonker}
{ G. H. Jonker and J. H. Van Santen, Physica {\bf 16}, 337 (1950);
J. H. Santen and G. H. Jonker, ibid. 599 (1950)}.

\bibitem{park}
{ J.-H. Park, et al., Phys. Rev. Lett. {\bf 76} 4215 (1996)}.

\bibitem{LaCa-end}
{ G. Matsumoto, J. Phys. Soc. Jpn. {\bf 29}, 606; 615 (1970)}.

\bibitem{bao95b}
{ A. P. Ramirez, et al.,
  Phys. Rev. Lett., {\bf 76} 3188 (1996)}.

\bibitem{bao94c}
{ P. Schiffer, A. P. Ramirez, W. Bao and S-W. Cheong, 
  Phys. Rev. Lett. {\bf 75} 3336, (1995)}.

\bibitem{theo}
{S. W. Lovesey, {\it Theory of Neutron Scattering from
  Condensed Matter}, (Clarendon Press, Oxford, 1984)}.


\bibitem{qzhu}
{Detail crystallographic study will be published separately.
 Q. Zhu {\it et al.}, (unpublished)}.

\bibitem{moonr}
{ R. M. Moon, T. Riste, W. C. Koehler, Phys. Rev. {\bf 181}, 920 (1969).
Finite counts for horizontal field with flipper off and vertical field with
flipper on are due to imperfection of the polarizing instruments}.

\bibitem{sw_La-Sr}
{ M. C. Martin, et al.,
  Phys. Rev. B {\bf 53}, 14285 (1996)}.

\bibitem{sw_La-Pb}
{ T. G. Perring, et al.,
Phys. Rev. Lett. {\bf 76}, 711 (1996)}.

\bibitem{sw_La-Ca}
{ J. W. Lynn, et al., Phys. Rev. Lett. {\bf 76}, 4046 (1996)}.

\bibitem{bao95a}
{ W. Bao, C. H. Chen, S. A. Carter and S-W. Cheong, 
  Solid State Comm. {\bf 98} 55, (1996)}.

\bibitem{Nd-Pb}
{ R. M. Kusters, et al., Physica {\bf B 155}, 362 (1989);
 K. Chahara, et al., Appl. Phys. Lett. {\bf 63}, 1990 (1993);
 H. L. Ju, et al., Phys. Rev. B {\bf 51}, 6143 (1995);
 A. Urushibara, et al., ibid. {\bf 51}, 14103 (1995);
 H. Y. Hwang, et al., Phys. Rev. Lett. {\bf 75}, 914 (1995)}.

\bibitem{two_l}
{ Y. Moritomo, et al., Nature {\bf 380}, 141 (1996)}.

\bibitem{hirota}
{ K. Hirota, et al., preprint (1996). Spin-wave excitations of A-type
antiferromagnetic LaMnO$_3$ was measured recently. $S J_F=-3.34(5)$meV,
$S J_{AF}=2.42(6)$meV and $H_u=5.3(1)$T.}


\end{references}
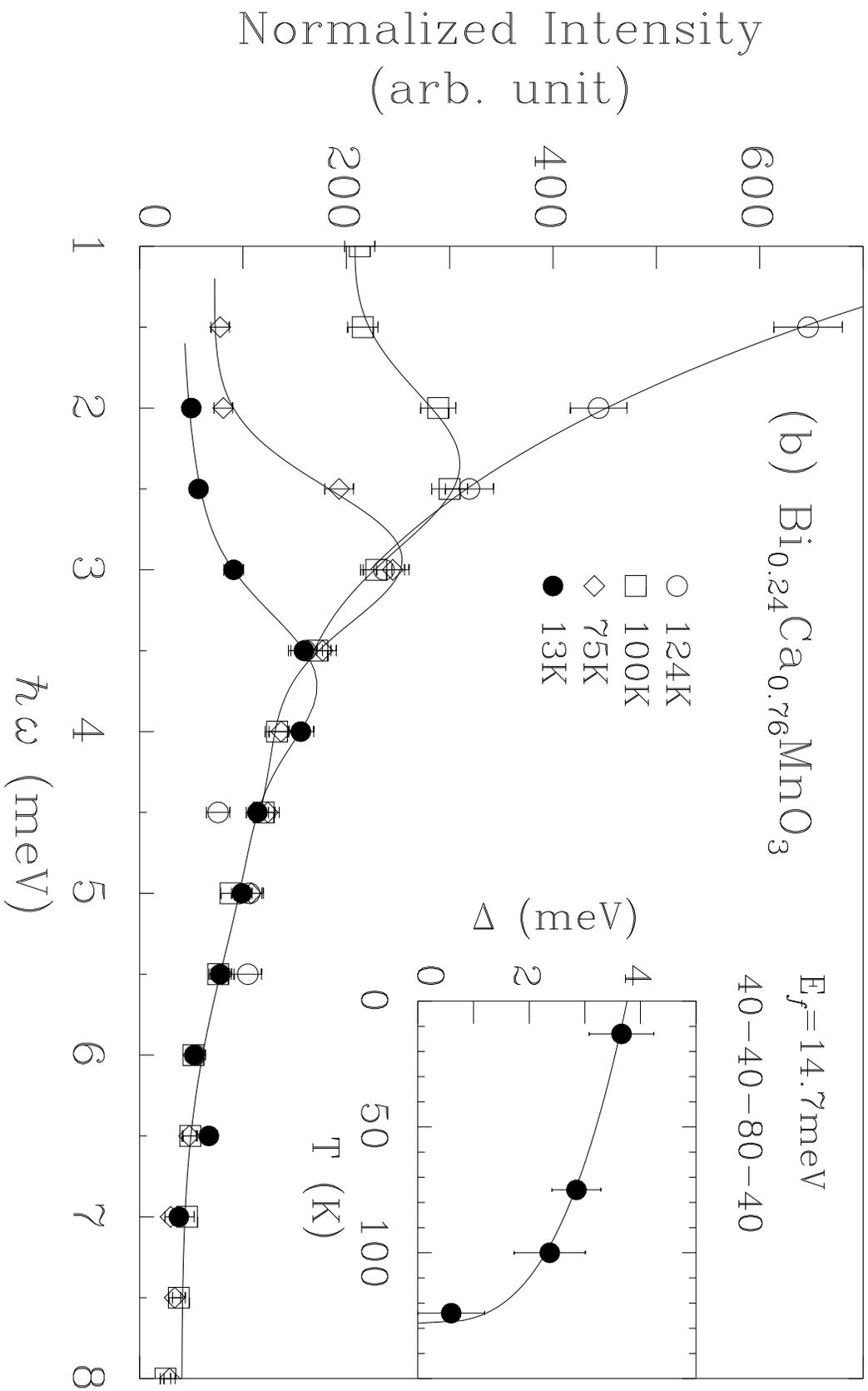
\end{document}